\documentclass{jaa}
\usepackage{natbib}
\usepackage{graphicx}

\begin{document}\sloppy

\title{Fractality of open clusters in singles, pairs, and groups}


\author{Almat Akhmetali\textsuperscript{1,2}}

\affilOne{\textsuperscript{1}Department of Electronics and Astrophysics, Al-Farabi Kazakh National University, 71 Al-Farabi Ave., Almaty, 050040, Kazakhstan\\}
\affilTwo{\textsuperscript{2}Institute of Experimental and Theoretical Physics, Al-Farabi Kazakh National University, 71 Al-Farabi Ave., Almaty, 050040, Kazakhstan}


\twocolumn[{

\maketitle

\corres{akhmetali\_almat@kaznu.edu.kz}

\msinfo{--}{--}

\begin{abstract}

In this work, we investigate the global structural properties and fractality of 1,876 open clusters (OCs) in different environments, including 1,145 singles, 392 pairs, and 339 groups. We analyze cluster mass, age, size, concentration, and fractal structure using the $Q$ parameter and the fractal dimension $f_{\mathrm{dim}}$, and examine their correlations with key physical parameters. Our results reveal systematic environmental trends: clusters in groups are generally younger, less massive, slightly larger, and less centrally concentrated than those in pairs or singles. Fractality is more pronounced in clusters within pairs and groups, with $44\%$ of group clusters exhibiting fractal substructure compared to $38.5\%$ for pairs and $33.2\%$ for singles. Similarly, median $f_{\mathrm{dim}}$ values increase from singles ($1.13$) to pairs ($1.16$) to groups ($1.25$), reflecting greater substructure in denser environments. These findings indicate that both intrinsic cluster properties and environmental context significantly influence cluster evolution. More massive clusters tend to evolve toward centrally concentrated, radially symmetric configurations, while less massive clusters retain fractal features for longer periods. Overall, our study demonstrates that OCs do not evolve in isolation: interactions with the environment play a critical role in shaping their structural evolution and dynamical state.

\end{abstract}

\keywords{Open clusters--fractality--fractal dimension.}
}]


\doinum{12.3456/s78910-011-012-3}
\artcitid{\#\#\#\#}
\volnum{000}
\year{0000}
\pgrange{1--}
\setcounter{page}{1}
\lp{1}

\section{Introduction}
\label{sec:Intro}
Open clusters (OCs) form in dense, turbulent molecular clouds from interstellar gas~\citep{lada2003}. After their formation, they can evolve into various configurations, including binary and higher multiplicity systems, offering a unique opportunity to study the processes of star formation and cluster evolution~\citep{camargo2016, shukirgaliyev2019star, shukirgaliyev2021bound, ishchenko2025dynamical, bissekenov2025evolution}. 

The dynamical evolution of clusters influences their spatial distribution and shapes their internal structure. Quantifying this spatial distribution provides valuable insights into their dynamical state and evolutionary history~\citep{ussipov2024}. In addition, the study of OC parameters such as mass, age, reddening, and metallicity allows detailed investigations that are not possible with field stars~\citep{piecka2021}.

According to the hierarchical star formation theory by~\cite{kruijssen2012}, young gravitationally bound OCs (ages $\leq100$ Myr) primarily form in high-density regions~\citep{vazquez2017, trevino2019, ward2020}. These clusters often show fractal substructures, consistent with the conveyor belt mechanism~\citep{clarke2010, arnold2017, fujii2022}. In contrast, clusters forming in low-density, filamentary environments typically exhibit filamentary substructures and tend to dissolve quickly following gas expulsion.

Internal two-body relaxation and external Galactic tidal forces play a significant role in the evolution of older OCs (ages $>100$ Myr). Two-body relaxation leads to the formation of a dense core surrounded by a low-density halo~\citep{pang2021}. Under strong Galactic tidal forces, extended tidal tails gradually develop over time~\citep{pang2022, tang2019, roser2019}.

The dynamical evolution of stellar clusters is a complex process governed by various factors. Both observational studies and numerical simulations suggest that clusters inherit a fractal spatial pattern from their parental molecular clouds~\citep{cartwright2004, clarke2010, sanchez2009, andre2010, andre2014, kuhn2014, jaehnig2015, arzoumanian2019, ballone2020}. With time, this initial fractality is progressively diminished by internal gravitational interactions and external perturbations. While unbound clusters disperse into the field, bound clusters contract toward their cores, developing more radially concentrated structures. However, this represents only a general evolutionary picture, as the timescale over which fractality is erased and the processes driving this transformation remain uncertain.

Interestingly, some very young clusters (e.g., $\rho$ Ophiuchus) already exhibit centrally concentrated morphologies at ages of only $\sim 1$ Myr, implying that dynamical evolution can proceed rapidly in certain environments~\citep{cartwright2004}. Conversely, pronounced substructure has been identified in much older clusters. For example, \cite{sanchez2009} reported significant subclustering in NGC~1513 and NGC~1641, both exceeding 100 Myr in age. In addition, recent $N$-body simulations demonstrate that, under specific conditions, clusters can develop centrally condensed distributions within only a few Myr~\citep{daffern2020}. Such contrasting findings underline the diversity of cluster evolutionary pathways and reinforce the need for further detailed investigations of their fractal structure.

In this work, we investigate the global and internal structural properties of OCs across different environments, including isolated clusters (singles), pairs, and groups. We analyze parameters such as age, mass, size, concentration, the $Q$ parameter, and the fractal dimension $f_{\mathrm{dim}}$ to assess how environmental complexity affects cluster characteristics. Our study examines both the overall distributions and the correlations of these parameters.

This paper is organized as follows. In Section~\ref{Sec:Sample}, we describe the sample of OCs used in this study. Section~\ref{Sec:Fractality} introduces the concept of fractality, with Section~\ref{sec:Q} detailing the computation of the $Q$ parameter and Section~\ref{sec:FD} describing the calculation of the fractal dimension $f_{\mathrm{dim}}$. In Section~\ref{Sec:Results}, we analyze the global properties of OCs in different environments (Section~\ref{Sec:Global}) and investigate the distributions and correlations of the $Q$ parameter (Section~\ref{Sec:Q}) and $f_{\mathrm{dim}}$ (Section~\ref{sec:FD}) with cluster properties. Finally, Section~\ref{sec:Conclusions} presents our conclusions and summarizes the main findings of this work.

\section{The sample of open clusters}
\label{Sec:Sample}
 
The OCs catalog used in this study is based on the dataset compiled by~\cite{palma2025}, which builds upon the work of~\cite{hunt2023, hunt2024}. In particular,~\cite{hunt2023} conducted a comprehensive all-sky search for OCs using Gaia DR3 data~\citep{vallenari2023}\footnote{https://www.cosmos.esa.int/web/gaia/dr3}, resulting in the largest and most homogeneous catalog of OCs to date. The catalog contains 7167 clusters, including 4782 previously known and 2387 newly identified ones. Cluster detection was performed using the Hierarchical Density-Based Spatial Clustering of Applications with Noise (HDBSCAN) algorithm, which is effective in identifying structures of varying density and separating them from background noise. In addition to cluster identification, the catalog provides key astrometric and astrophysical parameters, such as proper motions, parallaxes, ages, extinctions, distances, and photometric masses. The authors also evaluated the dynamical state of each system, distinguishing between bound clusters and unbound moving groups (MGs). They found that about 79\% of the systems in the catalog are consistent with being gravitationally bound OCs. This comprehensive dataset serves as the foundation for our analysis, allowing for a detailed study of the structural and dynamical properties of both known and newly discovered clusters.

~\cite{palma2025} calculated tidal forces that act on each cluster, taking into account only the influence of the nearest neighbour. The tidal force is estimated using the tidal factor, defined as \( \frac{d^3}{M_b \cdot r_{50}} \), where \( d \) is the distance to the nearest neighboring cluster (in parsecs), \( M_b \) is the total mass of that neighboring cluster (in solar masses), and \( r_{50} \) is the radius containing 50\% of the cluster members within the tidal radius (also in parsecs). This tidal factor is inversely related to the tidal force acting on the cluster. 

Based on the tidal forces of clusters acting each other, they classified clusters as following:

- Groups (G): Sets of three or more clusters, each located within 50 parsecs of the others, with all members having tidal factor values below 200.

- Pairs (P): Two neighboring clusters separated by less than 50 parsecs, where at least one has a tidal factor below 200, and no additional clusters are found nearby.

- Singles (S): Isolated clusters with no neighboring clusters within a 100-parsec radius.

- Unclassified: Clusters that do not meet the criteria for any of the categories above.

The study identified 2,052 isolated star clusters, 1,234 clusters in pairs, and 936 clusters grouped together. Pairs are further categorized as genetic binaries (B) if they formed simultaneously; tidal captures (C); or optical binaries (O), where the clusters appear close in projection but are not gravitationally bound. Optical binaries of the same age are labeled as Oa. Further information is available in~\cite{palma2025}.

In this study, we focus exclusively on the sample of OCs within the three possible environments: singles, pairs, and groups. We filtered sample clusters from the 7167 clusters by seven constraints: 1) “kind” = “o”; 2) S /Ns $>$ 5; 3) class\_50 (median CMD classifications) $>$  0.5; 4) membership probability $>$  0.5; 5) parallax\_error $<$  0.1 ; 6) photometric\_error $<$  0.1 (\citep{lindegren2018}, Appendix C); 7) members’ number (N) $>$  30 ; 8) G $>$ 18. The first three criteria help ensure that the selected objects are highly reliable OCs~\citep{hunt2023}. It is important to note that the membership probability threshold ($>$  0.5) effectively filters out low-quality members \cite{hunt2023}, which improves the reliability of the identified cluster members. Although this cut may exclude some stars located in tidal tails~\citep{hunt2023}, it does not remove all of them. As a result, some sample clusters may still contain both gravitationally bound and unbound stars.

Finally, we constrained our analysis to stars with a G magnitude brighter than 18 mag to avoid problems with the completeness of the sample at fainter magnitudes.

This selection results in a sample of 1145 single clusters, 392 clusters in pairs, and 339 clusters in groups, 1876 clusters in total. To achieve our objectives, we utilized the physical parameters derived for the full cluster sample by~\cite{hunt2023}, with a specific focus on total mass, size, and age.

\section{Fractality}
\label{Sec:Fractality}

In this work, the fractality of OCs was quantified using two complementary metrics: the $Q$ parameter and the fractal dimension, $f_{\mathrm{dim}}$. The $Q$ parameter provides a measure to distinguish between fractal and non-fractal distributions, while $f_{\mathrm{dim}}$ quantifies the complexity of the spatial arrangement of stars.

\subsection{Q parameter}
\label{sec:Q}

We estimated the $Q$ parameter, one of the most widely used measures of fractality, originally introduced by~\cite{cartwright2004}. It is commonly used in the analysis of both observational data and numerical simulations~\citep{schmeja2006, bastian2009, cartwright2009, sanchez2009, maschberger2010, parker2012, parker2014, parker2015, ballone2020, Laverde-Villarreal2025, coenda2025}.

The $Q$ parameter is defined as the ratio between the mean edge length of the minimum spanning tree (MST), $\bar{m}$, and the mean stellar separation, $\bar{s}$:
\begin{equation}
    Q = \frac{\bar{m}}{\bar{s}} .
\end{equation}
For consistency, $\bar{s}$ is normalized by a characteristic cluster radius $r_{\mathrm{cl}}$, defined as half of the maximum stellar separation, i.e., the radius of the smallest circle enclosing all members. Following~\cite{cartwright2009}, the MST mean edge length $\bar{m}$ is normalized by $(r_{\mathrm{cl}}^{2} N_s)^{1/2}$.

Low $Q$ values indicate that the MST edges are much shorter than the mean stellar separation ($\bar{m} \ll \bar{s}$), corresponding to a highly subclustered configuration of compact groups separated by wide gaps. In contrast, high $Q$ values correspond to smooth, centrally concentrated, or nearly uniform distributions where $\bar{m}$ and $\bar{s}$ are comparable. In practice, values of $Q < 0.8$ are typically associated with fractal or substructured morphologies, while $Q > 0.8$ indicates centrally concentrated structures. Values around $Q \approx 0.8$ correspond to approximately uniform stellar distributions.

\subsection{Fractal dimension}
\label{sec:FD}

We evaluated the structural complexity of OC distributions using the fractal dimension, which provides a quantitative measure of the spatial arrangement of stars. Lower fractal dimension values correspond to clumpy, substructured morphologies, while higher values indicate smoother and more centrally concentrated configurations~\citep{qin2025}.

In this work, we adopted the box-counting dimension, also known as the Minkowski–Bouligand dimension~\citep{imre2006}, estimated with the box-counting method of~\cite{grassberger1983}. The procedure consists of covering the dataset with boxes of varying sizes and counting how many are required to enclose all stars. Prior to applying this method, coordinates of the cluster members were standardized to zero mean and unit variance. This transformation renders the coordinates dimensionless, and consequently the box size $L$ used in the calculation is also dimensionless. Standardization was performed with the \texttt{StandardScaler} from the \texttt{scikit-learn} library~\citep{pedregosa2011scikit}.

\begin{figure}[h!]
\centering
\includegraphics[width=0.95\linewidth]{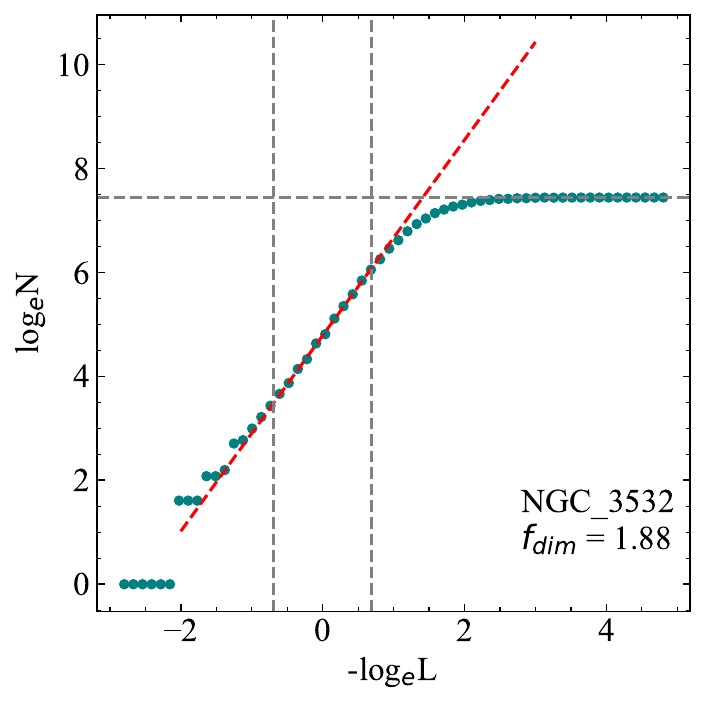}
  \caption{Box-counting  $\log_e N(L)$ vs. $-\log_e L$ for the NGC 3532 cluster. The red solid line represent the best-fit robust linear regression, with its slope corresponding to the estimated fractal dimension $f_\mathrm{dim}$. The vertical dashed lines mark the fitting range of $[-\log_e 2, \log_e 2]$. The horizontal dashed lines indicate the natural logarithm of the total number of member stars in the cluster.}
     \label{fig:FD_example}
\end{figure}

The fractal dimension, $f_{\mathrm{dim}}$, is defined as
\begin{equation}
    f_{\mathrm{dim}} = -\frac{d \log N(L)}{d \log L},
\end{equation}
where $L$ denotes the box size and $N(L)$ the number of boxes required to cover the stellar distribution. Since we deal with a finite, discrete set of points, the derivative is approximated using finite differences. In practice, $f_{\mathrm{dim}}$ is estimated as the slope of the linear regression of $\log N(L)$ versus $-\log L$ (see Fig.~\ref{fig:FD_example}).

As illustrated in Fig.~\ref{fig:FD_example}, plateaus emerge when $N(L)$ approaches the total number of stars (horizontal dashed line). Following~\cite{qin2025}, we restricted the fitting range to $[-\log 2, \log 2]$, indicated by vertical dashed lines, to mitigate biases introduced by these plateaus and to ensure consistency. This interval corresponds to box sizes between half and twice the characteristic scale set by the coordinate standard deviation, which is roughly equivalent to the half-mass radius of the cluster~\citep{qin2025}.

\subsection{Comparison of methods}
\label{sec:Comparison}

The use of both the $Q$ parameter and the fractal dimension in this work is motivated by their complementary capabilities in characterizing the spatial morphology of OCs. Although both metrics aim to quantify structural complexity, they are derived from different conceptual frameworks and respond to different aspects of the stellar distribution.

The $Q$ parameter provides a global measure of structure by comparing the mean edge length of the minimum spanning tree with the mean stellar separation (see Section~\ref{sec:Q}). It is primarily sensitive to the relative dominance of small-versus large-scale clustering and offers a convenient threshold-based classification of stellar configurations. Specifically, it allows the distinction between substructured (fractal-like) and centrally concentrated morphologies through a single scalar value. Due to its computational simplicity and robustness against variations in cluster size and density, it is frequently used as a first-order diagnostic in both observational and numerical studies~\citep{sanchez2009,parker2015,Laverde-Villarreal2025, coenda2025}.

On the other hand, the fractal dimension offers a more detailed quantification of spatial complexity by evaluating how the number of occupied regions scales with observational resolution. This method is inherently sensitive to scale-dependent features and captures hierarchical substructure that may not significantly influence global metrics like the $Q$ parameter. As such, the fractal dimension is particularly valuable for analyzing clusters formed through processes involving turbulent fragmentation or hierarchical assembly, where multi-scale structure is expected.

Despite their conceptual similarities, the two methods do not always yield identical classifications. For instance, clusters with $Q$ values near the transitional threshold ($Q \approx 0.8$) may exhibit a broad range of fractal dimensions, indicating underlying complexity that is not reflected in the $Q$ parameter alone. Conversely, clusters with similar fractal dimensions may display different $Q$ values depending on their degree of central concentration or the presence of large-scale gradients.

The combined use of both methods thus enables a more comprehensive structural analysis. While the $Q$ parameter serves as a robust, model-independent classifier of gross morphological trends, the fractal dimension provides a scale-sensitive measure of internal substructure. Their combined application facilitates cross-validation of results and enhances the interpretability of structural diagnostics in the context of cluster formation and dynamical evolution.

\section{Results}
\label{Sec:Results}

In this section, we present the main findings of our analysis of OCs. We first examine the global properties of clusters across different environments, including age, mass, size, and concentration. We then investigate cluster substructure using both the $Q$ parameter and the fractal dimension $f_{\mathrm{dim}}$, analyzing how these measures correlate with global cluster properties. Throughout, we compare trends and differences among clusters classified as Singles, Pairs, and Groups, highlighting systematic patterns in structure and morphology.

\subsection{Global properties of open clusters in different environments}
\label{Sec:Global}

\begin{figure*}[h!]
\centering
\includegraphics[width=0.75\textwidth]{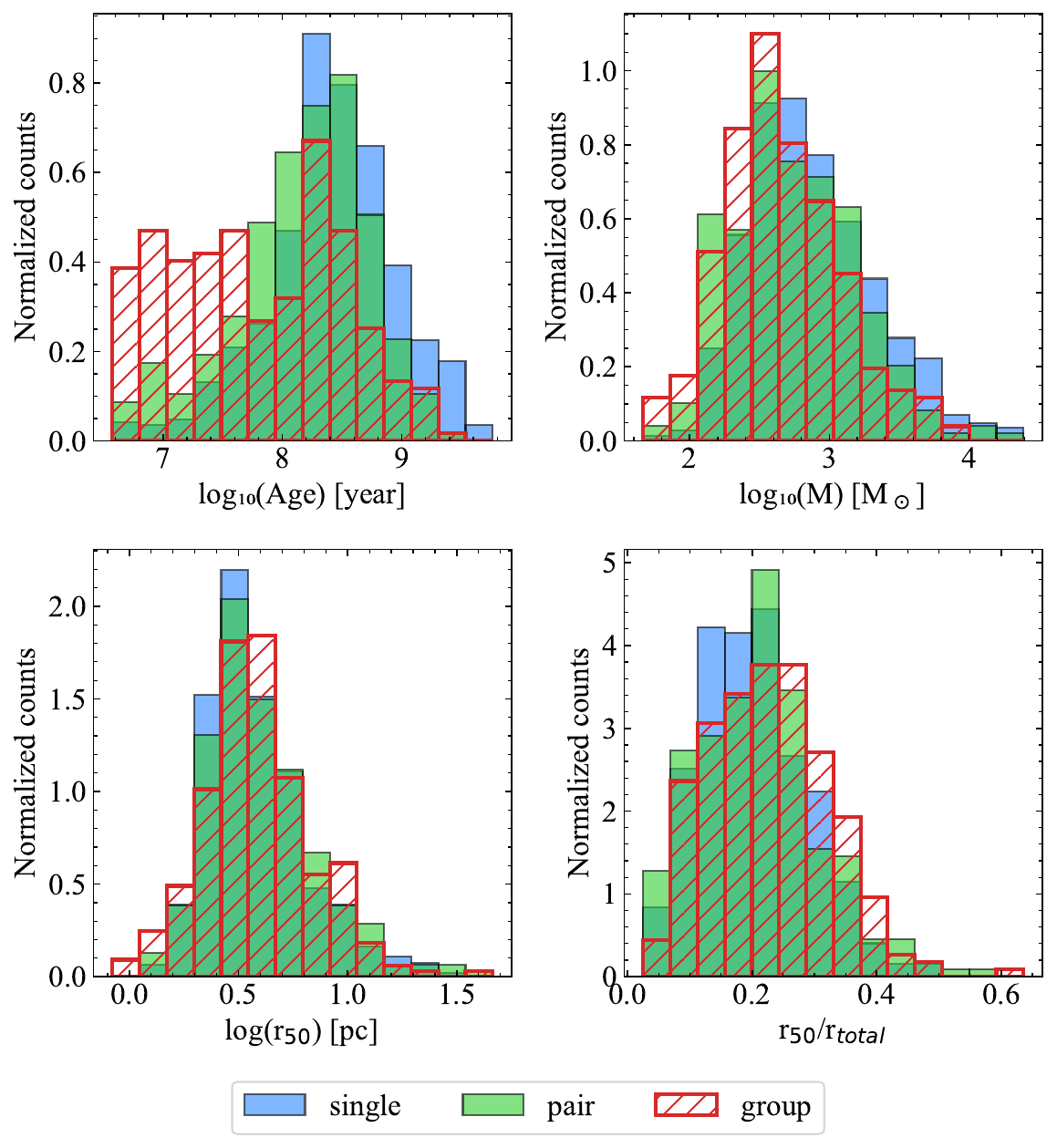}
  \caption{Global properties of the OCs sample. Singles are shown in blue, pairs in green, and groups in red.}
     \label{fig:Global_histograms}
\end{figure*}

To assess the impact of environmental conditions on OC characteristics, we first examine their global properties across different environments. Figure~\ref{fig:Global_histograms} presents the distributions of key parameters (age, mass, size, and concentration) for our OC sample, while Table~\ref{tab:table1} provides a summary of statistics, including median values, bootstrap uncertainties, and $95\%$ confidence intervals.

\begin{table}[ht]
\centering
\setlength{\tabcolsep}{2pt} 
\caption{Summary statistics for various parameters of OCs, categorized as Singles, Pairs, and Groups. For each parameter and cluster type, we report the median with bootstrap error, and the 95\% confidence interval (CI).
Mass, Age, and $r_{50}$ are given in $\log_{10}$ units.}
\label{tab:table1}
\begin{tabular}{lcccc}
\hline
\textbf{Parameter} & \textbf{Type} & \textbf{Median $\pm$ Error} & \textbf{95\% CI} \\
\hline
$\log_{10}$(Age) [year] & Single & $8.40 \pm 0.03$ & $8.37 - 8.43$ \\
                  & Pair   & $8.22 \pm 0.07$ & $8.16 - 8.29$ \\
                  & Group  & $7.77 \pm 0.13$ & $7.61 - 7.86$ \\
\hline
$\log_{10}$(M) [M$_\odot$] & Single & $2.71 \pm 0.03$ & $2.68 - 2.73$ \\
                  & Pair   & $2.61 \pm 0.04$ & $2.57 - 2.65$ \\
                  & Group  & $2.55 \pm 0.04$ & $2.50 - 2.58$ \\                  
\hline
$\log_{10}(r_{50})$ [pc] & Single & $0.53 \pm 0.01$ & $0.51 - 0.54$ \\
                    & Pair   & $0.55 \pm 0.03$ & $0.52 - 0.58$ \\
                    & Group  & $0.56 \pm 0.02$ & $0.53 - 0.58$ \\
\hline
$r_{50}/r_{\mathrm{total}}$ & Single & $0.20 \pm 0.01$ & $0.20 - 0.21$ \\
                            & Pair   & $0.22 \pm 0.01$ & $0.21 - 0.24$ \\
                            & Group  & $0.23 \pm 0.01$ & $0.22 - 0.24$ \\
\hline
$Q$ parameter & Single & $0.87 \pm 0.01$ & $0.86 - 0.88$ \\
              & Pair   & $0.84 \pm 0.02$ & $0.82 - 0.86$ \\
              & Group  & $0.82 \pm 0.01$ & $0.81 - 0.84$ \\
\hline
$f_{\mathrm{dim}}$ & Single & $1.13 \pm 0.02$ & $1.11 - 1.15$ \\
                   & Pair   & $1.16 \pm 0.03$ & $1.12 - 1.19$ \\
                   & Group  & $1.25 \pm 0.03$ & $1.23 - 1.29$ \\
\hline
\end{tabular}
\end{table}

Clusters in the most complex environments (groups) are the youngest, with a median age of $(7.77 \pm 0.13)$, followed by pairs $(8.22 \pm 0.07)$, and singles $(8.40 \pm 0.03)$, indicating that more isolated clusters tend to be older. A similar but weaker trend is observed for masses: groups exhibit the lowest median mass $(2.55 \pm 0.04)$, pairs are intermediate $(2.61 \pm 0.04)$, and singles are the most massive $(2.71 \pm 0.03)$. Cluster sizes, measured by $r_{50}$, follow the same ordering, with groups slightly larger $(0.56 \pm 0.02)$, pairs $(0.55 \pm 0.03)$, and singles $(0.53 \pm 0.01)$. Finally, the concentration parameter $r_{50}/r_{\mathrm{total}}$ reflects this trend: groups are the least concentrated $(0.23 \pm 0.01)$, pairs are intermediate $(0.22 \pm 0.01)$, and singles are the most concentrated $(0.20 \pm 0.01)$.

\begin{figure*}[h!]
\centering
\includegraphics[width=1\textwidth]{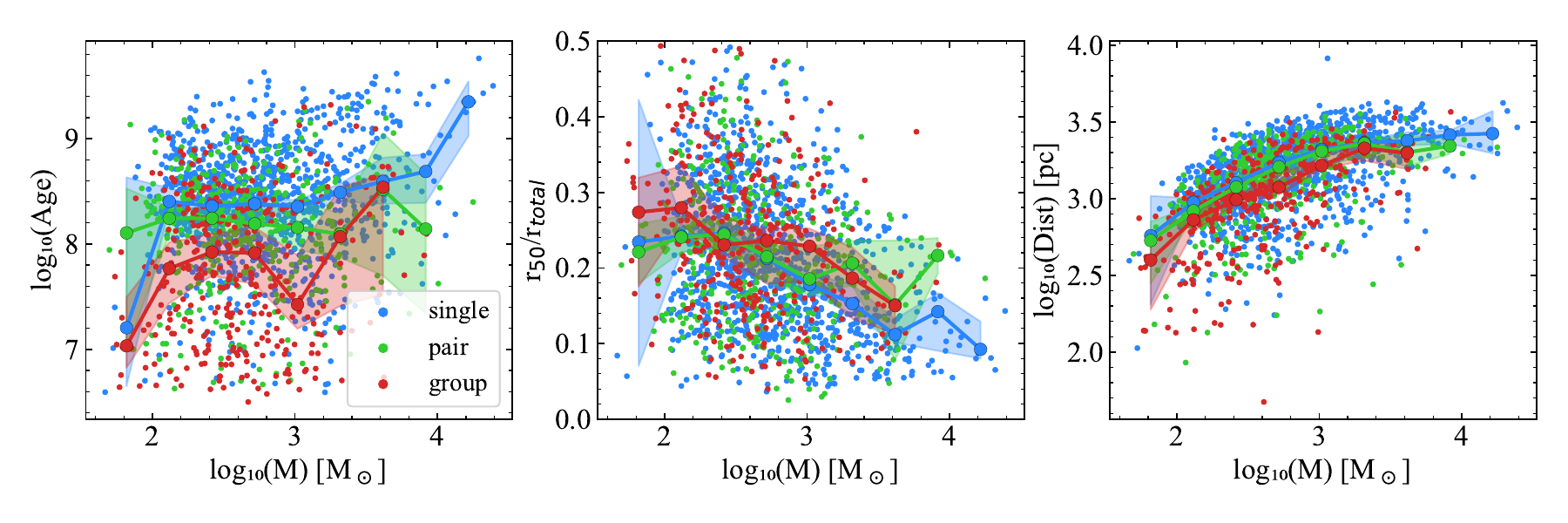}
    \caption{Median values of global parameters (from left to right: age, concentration and distance) as a function of cluster mass for our sample of OCs. Shaded areas indicate uncertainties estimated using the bootstrap method. Singles are shown in blue, pairs in green, and groups in red.}
        \label{fig:Global_properties}
 \end{figure*}

Figure~\ref{fig:Global_properties} illustrates the median values of cluster parameters—age, concentration, and distance—as a function of cluster mass. Shaded regions represent uncertainties estimated via bootstrap resampling. The left panel shows that median cluster age increases with mass across singles, pairs, and groups. The middle panel exhibits a negative correlation between cluster concentration and mass. The right panel demonstrates a steady increase in cluster distance with mass, with relatively low uncertainty.

Overall, these results indicate a consistent pattern in which clusters residing in more complex environments tend to be younger, less massive, slightly larger, and more diffuse, whereas clusters in more isolated regions are older, more massive, smaller, and more centrally concentrated.

\subsection{Fractality of open clusters in different environments}
\label{Sec:Q}

\begin{figure}[h!]
\centering
\includegraphics[width=0.95\linewidth]{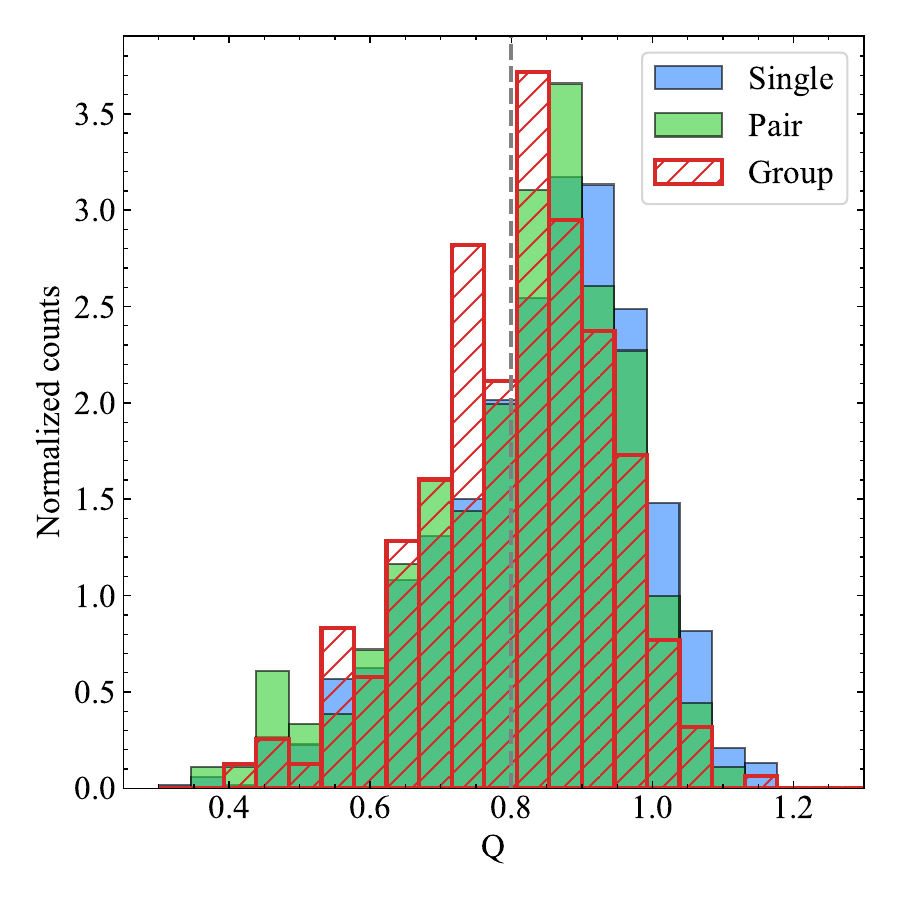}
  \caption{Histogram of the $Q$ parameter. Singles are shown in blue, pairs in green, and groups in red. The vertical grey dashed line marks $Q=0.8$, which serves as the threshold distinguishing fractal substructure from radial density profiles.}
     \label{fig:Q_histogram}
\end{figure}

The distribution of the $Q$ parameter for OCs in different environments is shown in Figure~\ref{fig:Q_histogram}. The vertical line at $Q = 0.8$ marks the threshold separating fractal substructures from radial density profiles (see Section~\ref{sec:Q}). As reported in Table~\ref{tab:table1}, the median $Q$ values are $0.87 \pm 0.01$ for singles, $0.84 \pm 0.02$ for pairs, and $0.82 \pm 0.01$ for groups. Correspondingly, the fractions of clusters exhibiting fractal structure are $33.2\%$ for singles, $38.5\%$ for pairs, and $44.0\%$ for groups. These results suggest that clusters in more complex environments are more likely to exhibit fractal substructures than isolated clusters. This finding is consistent with the idea that clusters inherit fractality from their parental molecular clouds. Over time, this fractal structure tends to diminish, which aligns with the observation that clusters in groups are systematically younger than pairs or singles (see Figure~\ref{fig:Global_histograms}).

\begin{figure*}[h!]
\centering
\includegraphics[width=0.75\textwidth]{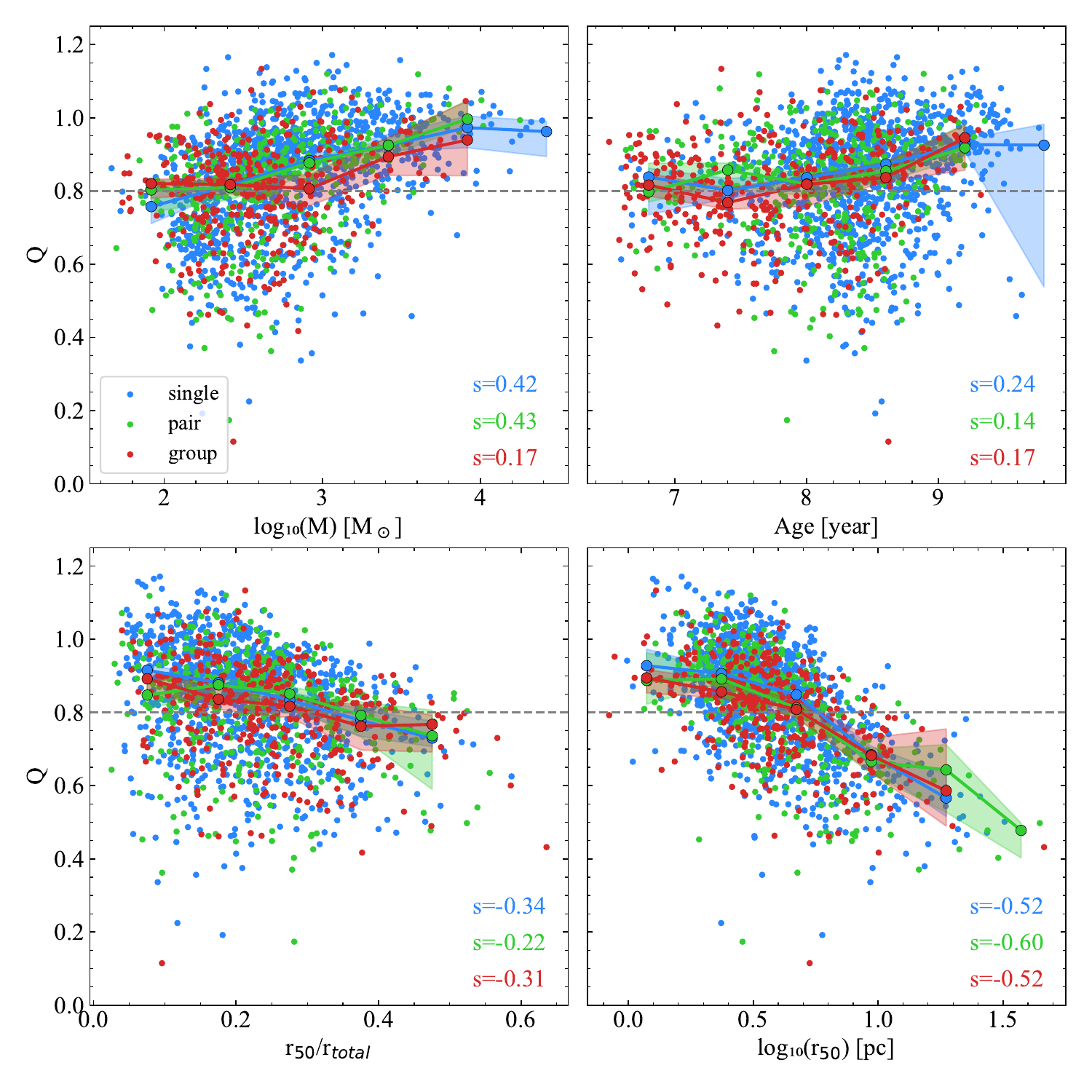}
    \caption{$Q$ parameter as a function of various global cluster parameters, including mass, age, concentration, and cluster size. Singles are shown in blue, pairs in green, and groups in red. Shaded areas indicate uncertainties estimated using the bootstrap method. The Spearman correlation coefficient for each panel is indicated in the lower-right corner, with colors corresponding to the cluster types. Positive values of $s$ indicate a positive correlation, with a maximum of 1 representing a perfect positive correlation. Negative values indicate a negative correlation, while 0 corresponds to no correlation. The horizontal grey dashed lines marks $Q=0.8$, which serves as the threshold distinguishing fractal substructure from radial density profiles.}
     \label{fig:Q}
\end{figure*}

Figure~\ref{fig:Q} shows the $Q$ parameter as a function of global cluster properties, including mass, age, concentration, and size. The Spearman correlation coefficient~\citep{zwillinger1999}, denoted as $s$, is reported in the lower-right corner of each panel, with colors corresponding to different cluster types. Positive values of $s$ indicate a positive correlation, with a maximum of 1 representing a perfect positive correlation; negative values indicate a negative correlation, and 0 indicates no correlation. In the top-left panel, a moderate correlation is observed between $Q$ and mass for pairs, while singles and groups show weaker correlations. The top-right panel reveals only weak correlations with age. Similarly, weak correlations are seen with cluster concentration in the bottom-left panel, whereas the bottom-right panel shows a moderate correlation between $Q$ and cluster size.

\subsection{Fractal dimension of open clusters in different environments}
\label{Sec:FD}

\begin{figure}[h!]
\centering
\includegraphics[width=0.95\linewidth]{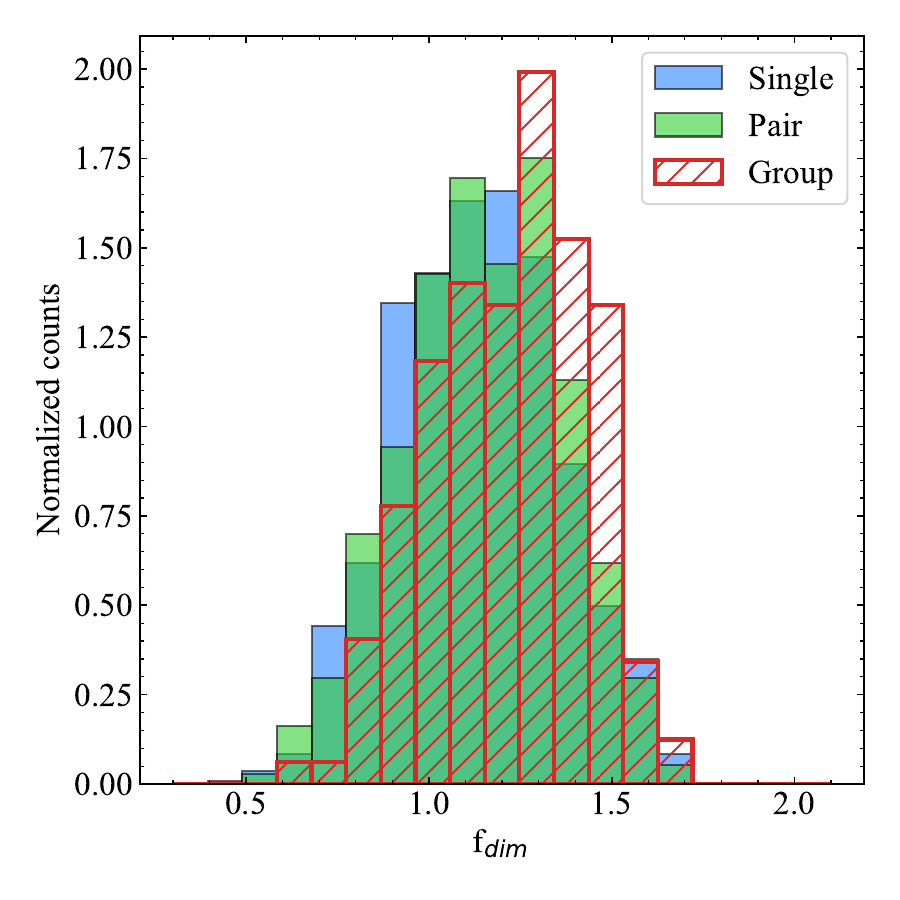}
  \caption{Histogram of the $f_{\mathrm{dim}}$. Singles are shown in blue, pairs in green, and groups in red.}
     \label{fig:FD_histogram}
\end{figure}

The distribution of $f_{\mathrm{dim}}$ for OCs in different environments is shown in Figure~\ref{fig:FD_histogram}. As reported in Table~\ref{tab:table1}, the median $f_{\mathrm{dim}}$ values are $1.13 \pm 0.02$ for singles, $1.16 \pm 0.03$ for pairs, and $1.25 \pm 0.03$ for groups. This suggests that clusters in more complex environments (groups) tend to retain a higher degree of fractal substructure, whereas clusters in more isolated environments (singles) are more centrally concentrated and dynamically relaxed, as lower $f_{\mathrm{dim}}$ values correspond to more substructured distributions (\ref{sec:FD}). The gradual increase of $f_{\mathrm{dim}}$ from singles to pairs to groups aligns with the trend observed for the $Q$ parameter, further supporting the idea that environmental complexity influences the internal spatial structure of OCs.

\begin{figure*}[h!]
\centering
\includegraphics[width=0.75\textwidth]{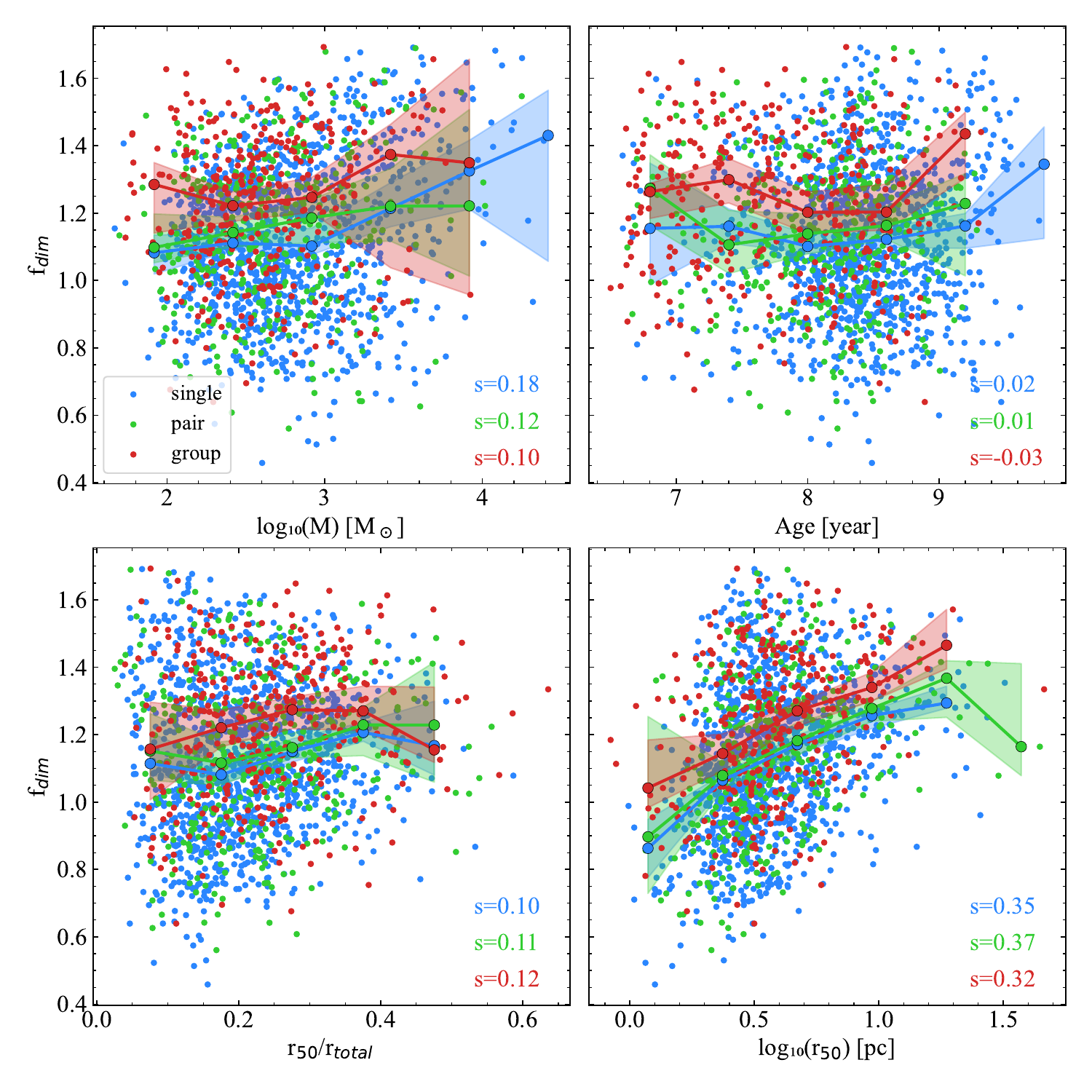}
  \caption{$f_{\mathrm{dim}}$ as a function of various global parameters, including mass, age, concentration, and cluster size. Singles are shown in blue, pairs in green, and groups in red. Shaded areas indicate uncertainties estimated using the bootstrap method. The Spearman correlation coefficient for each panel is indicated in the lower-right corner, with colors corresponding to the cluster types. Positive values of $s$ indicate a positive correlation, with a maximum of 1 representing a perfect positive correlation. Negative values indicate a negative correlation, while 0 corresponds to no correlation.}
  
     \label{fig:FD}
\end{figure*}

Figure~\ref{fig:FD} shows $f_{\mathrm{dim}}$ as a function of global cluster properties, including mass, age, concentration, and size. The Spearman correlation coefficient, denoted as $s$, is indicated in the lower-right corner of each panel, with colors corresponding to different cluster types. We find no significant correlation between $f_{\mathrm{dim}}$ and cluster mass, age, or concentration. In contrast, a moderate correlation is observed with cluster size, which is expected, as cluster size directly influences the spatial distribution and, consequently, the measured $f_{\mathrm{dim}}$.

Despite both the $Q$ parameter and $f_{\mathrm{dim}}$ being used to quantify fractality, they show opposite trends with cluster size in Figures~\ref{fig:Q} and~\ref{fig:FD}. The $Q$ parameter decreases as cluster size increases, indicating that larger clusters appear more substructured, whereas the $f_{\mathrm{dim}}$ increases with cluster size, suggesting larger clusters are smoother. This difference likely stems from the distinct aspects of spatial structure that each method captures. The $Q$ parameter provides a global measure that emphasizes overall clustering patterns, while the $f_{\mathrm{dim}}$ focuses on how complexity changes across scales and is more sensitive to small-scale features and cluster boundaries. Additionally, the $f_{\mathrm{dim}}$ results exhibit greater scatter, which may be due to its sensitivity to sample size and the specific box-counting procedure. These findings imply that while both metrics are useful, they reflect different structural properties, and should be interpreted together to gain a more comprehensive understanding of cluster morphology.

\section{Discussion}
\label{sec:Discussion}

Our analysis reveals clear differences in the global properties and fractal structures of OCs depending on their environmental context. Clusters residing in groups tend to be younger, less massive, larger, and exhibit more pronounced fractal substructures than those found in pairs or isolated singles. This suggests that environmental complexity plays a critical role in shaping cluster properties and their internal spatial morphology.

These findings align well with current theories of star formation within turbulent, hierarchical molecular clouds. Star formation occurs in clouds with complex, filamentary density structures, leading to the formation of substructured stellar groupings. The fractal substructures observed in clusters, particularly those in groups, likely reflect this imprint of the parental cloud. Turbulence generates regions of varying density and fragmentation scales, leading to hierarchical clustering of newly formed stars. Because clusters in groups are systematically younger, they retain more of this initial fractal pattern and have undergone less dynamical evolution.

In contrast, single clusters tend to be older, more massive, smaller, and more centrally concentrated, consistent with a more dynamically evolved state. Over time, processes such as two-body relaxation, tidal interactions, and gas expulsion act to erase the initial fractal substructure, yielding a smoother and more radially symmetric spatial distribution. The dispersal of neighboring groups and dissolution of the molecular cloud further contributes to their apparent isolation.

The observation that isolated clusters tend to be older can be interpreted as an evolutionary result rather than merely a reflection of initial conditions. By definition, isolated clusters in our sample have no neighbors within 100 parsecs. As clusters age, their parental molecular clouds dissolve, and the nearby low-mass stellar groups formed within the same cloud also disperse or disrupt. Consequently, the initially clustered environment becomes less crowded, leaving older clusters appearing more isolated. This evolutionary scenario naturally explains the systematic age difference observed between isolated clusters and those in groups.

Overall, our results support a scenario in which the fractal structure of young clusters is a direct inheritance from the turbulent and hierarchical nature of their natal molecular clouds. Environmental factors, including cluster grouping and local stellar density, influence both initial cluster formation and subsequent dynamical evolution, shaping observable properties such as age, mass, concentration, and fractality. Future studies incorporating kinematic data and numerical simulations of cluster formation and evolution will be instrumental in deepening our understanding of these processes and the role of environment in star cluster development.

\section{Conclusion}
\label{sec:Conclusions}

In this paper, we investigated the global properties of OCs, including mass, age, size, and fractality, across different environments. Our sample comprises a total of 1,876 clusters, consisting of 1,145 singles, 392 pairs, and 339 groups. To assess fractality, we computed the $Q$ parameter and the fractal dimension $f_{\mathrm{dim}}$, and examined their correlations with key cluster parameters such as mass and age. The main findings are as follows:

-- Clusters in groups are, on average, younger, less massive, slightly larger, and more diffuse, followed by clusters in pairs and single clusters.

-- Clusters in groups exhibit fractal structures more frequently (44\%) compared to clusters in pairs (38.5\%) and singles (33.2\%).

-- Clusters in groups have higher median values of $f_{\mathrm{dim}}$ (1.25) than pairs (1.16) and singles (1.13).

These results demonstrate that both the intrinsic properties of clusters and their environmental context strongly influence their evolution. The higher fractality observed in clusters within pairs and groups indicates that their structural development remains closely tied to the initial substructure and dynamics of their parental molecular clouds. While more massive clusters tend to evolve toward centrally concentrated, radially symmetric configurations, less massive clusters are able to retain their fractal substructure for longer periods. Overall, our findings emphasize that OCs do not evolve in isolation: single clusters exhibit clear signs of advanced dynamical relaxation, whereas clusters in pairs and groups preserve structural imprints of their formation environment, highlighting the crucial role of environmental interactions in shaping their long-term evolution.

\section*{Acknowledgements}
We thank Dr Nurzhan Ussipov and Dr Bekdaulet Shukirgaliyev for their useful discussions about fractality and open clusters.

\bibliography{main}

\end{document}